\documentclass[epj]{mysvjour}
\usepackage{graphics}
\begin{document}
\title{Moments of Isovector Quark Distributions in Lattice QCD}
\author{W. Detmold\inst{1}\inst{2} \and W.~Melnitchouk\inst{3} \and
  A.W.~Thomas\inst{2}
}                     
\institute{Department of Physics, University of Washington, Box
  351560, Seattle, WA 98195, U.S.A. \and Special Research Centre for
  the Subatomic Structure of Matter, University of Adelaide, Adelaide,
  SA 5005, Australia. \and Jefferson Lab, 12000 Jefferson Avenue,
  Newport News, VA 23606, U.S.A.}
\date{Received: date / Revised version: date}
%
\abstract{ We investigate the connection of lattice calculations of
  moments of isovector parton distributions to the physical regime
  through extrapolations in the quark mass. We consider the one pion
  loop renormalisation of the nucleon matrix elements of the
  corresponding operators and thereby develop formulae with which to
  extrapolate the moments of the unpolarised, helicity and
  transversity distributions. These formulae are consistent with
  chiral perturbation theory in the chiral limit and incorporate the
  correct heavy quark limits. In the polarised cases, the inclusion of
  intermediate states involving the $\Delta$ isobar is found to be
  very important.  The results of our extrapolations are in general
  agreement with the phenomenological values of these moments where
  they are known, and for the first time we perform an extrapolation
  of the low moments of the isovector transversity distribution which
  is consistent with chiral symmetry.
} 
\maketitle
\section{Introduction}
\label{intro}

Until recently, state of the art studies \cite{latt} of the moments of
parton distributions within lattice QCD have led to large
discrepancies with experimental data \cite{param}. Lattice calculations
are performed at quark masses much greater than those of the physical
light quarks. Consequently, an extrapolation to the physical mass
regime must be made in order to compare with experiment. A naive
linear extrapolation results in values for the first three non-trivial
moments of the unpolarised $u-d$ distribution that are 50\% above the
phenomenological values \cite{param}. For the polarised distributions
the results are no better; $g_A$ (the 0$^{\rm th}$ moment of $\Delta
u- \Delta d$) is significantly underestimated by a linear
extrapolation, while the lack of accuracy in the data (both
experimental and lattice) on higher moments precludes definitive
statements. Such a large disagreement in such basic hadronic
observables casts doubt on the current reliability of the lattice
approach to hadronic physics.

We present an improved extrapolation scheme
\cite{DMNRT,DMT,DMT2,WDThesis} that for the first time resolves much
of this discrepancy. Using constraints from chiral symmetry and the
heavy quark limit, we develop a formalism for the extrapolation of the
moments of the isovector, unpolarised, helicity and transveristy
distributions. These are related to the forward nucleon matrix
elements of various twist-2 operators which are calculated on the
lattice through the operator product expansion. In particular we find
that contributions from intermediate states involving the $\Delta$
isobar are large, and their effects cannot be ignored in any
quantitative analysis. With their inclusion, we are able to make
reliable, almost model-independent\footnote{Results are independent of
  the shape of the pion-nucleon form factor to 1-2\% \cite{Ross}.}
extrapolations of the moments of parton distributions and make
predictions for the low spin transversity moments which can be tested
by future measurements.

\section{Pion-Loop Renormalisation of Moments of Parton Distribution 
  Functions} 
General constraints from the approximate chiral symmetry
of QCD lead to the appearance of non-analytic terms in the quark mass
expansion of many hadronic quantities. In particular, the moments of
quark distributions behave as \cite{TMS,CHPT}
\begin{equation}
\langle x^n \rangle_{q,\Delta q,\delta q} \sim m_\pi^2\log m_\pi^2,
\end{equation}
where the Gell-Mann--Oakes--Renner relation, $m_q\sim m_\pi^2$, has
been used to express the quark mass in terms of the pion mass.  This
behaviour arises from the infrared properties of the one-pion loop
renormalisation of the matrix elements of the corresponding twist-2
operators ${\cal O}_{q,\Delta q,\delta q}$ (see Ref.~\cite{DMT2} for
their definitions and further details). That is,
\begin{equation}
  \label{merenorm}
  \langle N|{\cal O}^{\mu_1\ldots\mu_n}_i
  |N\rangle_{\rm dressed}
  = \frac{Z_2}{Z_i}
  \langle N|{\cal O}^{\mu_1\ldots\mu_n}_i
  |N\rangle_{\rm bare}\,,
\end{equation}
where $Z_2$ is the nucleon wavefunction renormalisation and $Z_i$, $i =
q,\Delta q,\delta q$, are the operator renormalisations arising from
$\pi N$, $\pi\Delta$, and $\pi(N\Delta$ transition) intermediate
states.

In the case of the unpolarised moments, simple one-loop calculations
of the renormalisation using a variety of form factors (cutoff,
monopole, dipole) for the pion--nucleon coupling all lead to a simple
extrapolation formula
\begin{equation}
  \label{xtrap}
       \langle x^n\rangle_{u-d} = a_n \left( 1 + c_{\rm
       LNA}m_\pi^2\log\frac{m_\pi^2}{m_\pi^2+\mu^2} \right) +
       b_n m_\pi^2\,,
\end{equation}
with three free parameters, $a_n$, $b_n$, $\mu$ ($c_{\rm LNA}$ is
fixed by chiral perturbation theory). Similar expressions are found
for the helicity and transversity moments in Ref.~\cite{DMT2}. The
parameter $\mu$ describes the physical scale of the pion source
\cite{Detmold:2001hq} and ideally would be constrained by lattice
data. However, until sufficiently accurate lattice data at low quark
masses are available, alternative methods (such as the one used here)
must be used to fix $\mu$.

\begin{figure}
  \resizebox{0.95\columnwidth}{!}{%
    \includegraphics{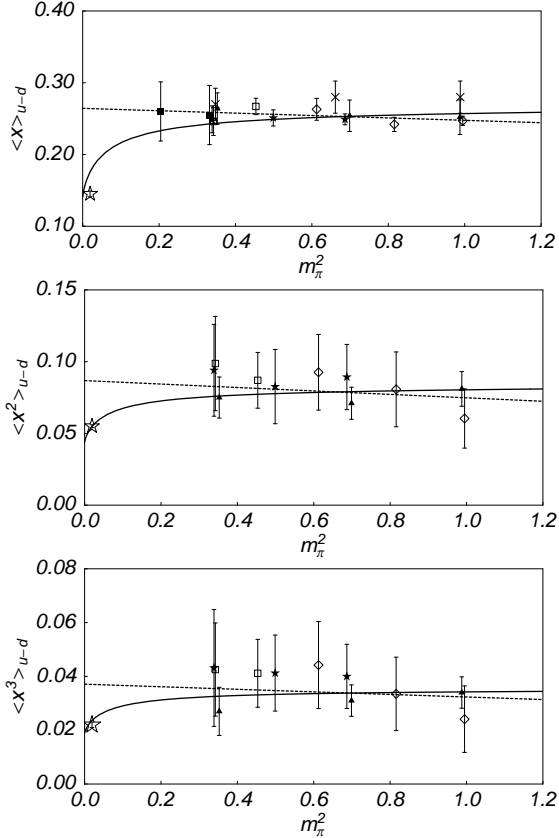}
    }
  \caption{\label{f1}Lowest three nontrivial moments of the
    unpolarised isovector parton distribution. Data are taken from
    various lattice simulations (see Ref.~\cite{DMT2} for details).
    The linear extrapolation (dashed) significantly overestimates the
    experimental results (stars), whilst the extrapolation using
    Eq.~(\ref{xtrap}) is in reasonable agreement.}
\end{figure}

\begin{figure}
  \resizebox{0.95\columnwidth}{!}{%
    \includegraphics{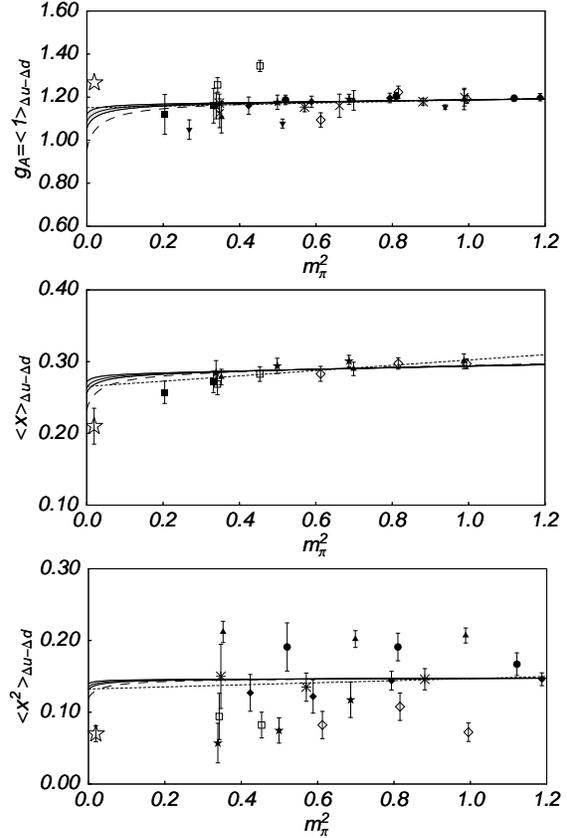}
    }
  \caption{\label{f2}Lowest three nontrivial moments of the helicity  
    isovector parton distribution. Curves shown are (dotted line)
    linear extrapolation, (dashed line) extrapolation using the
    polarised analog of Eq.~(\ref{xtrap}) but ignoring $\Delta$
    contributions, and (solid lines) extrapolations for varying values
    of the $\pi N\Delta$ coupling. (Data as for Fig.~\ref{f1}.)}
\end{figure}

As shown by the solid curves in Fig.~\ref{f1} this formula provides an
excellent description of the lattice data, the experimental values of
the moments and (with suitable modification -- see Ref.~\cite{DMT})
the values known in the heavy quark limit. However, when a similar
formula is applied to the moments of the helicity distributions, the
extrapolation is considerably worse than a naive linear extrapolation,
as shown by the dashed lines in Fig.~\ref{f2}.
 
\begin{figure}
  \resizebox{0.95\columnwidth}{!}{%
    \includegraphics{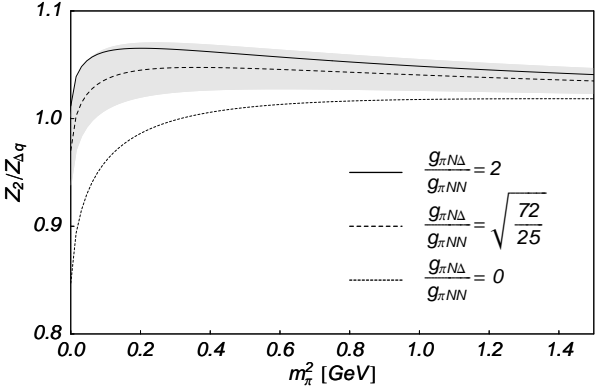}
    }
  \caption{\label{f3} One-loop renormalisation of the nucleon matrix
    elements of the polarised, isovector, twist-2 operators using a
    dipole form factor ($\Lambda=0.8$~GeV) and varying values of the
    coupling ratio $g_{\pi N\Delta}/g_{\pi NN}$. The shaded region
    corresponds to variation of the Weinberg-Tomozawa coupling -- see
    Ref.~\cite{DMT2}.}
\end{figure}

When we turn to the moments of polarised parton distributions, there
is considerable evidence from phenomenological models that suggests
the $\Delta$ resonance will play an important role.  Although the
$\Delta$ contributions formally enter at higher order in $m_\pi$, the
coefficients of these next-to-leading non-analytic terms are large,
and they cannot be ignored in any quantitative analysis.  This is
clearly demonstrated by Fig.~\ref{f3} where the pion mass dependence
of the one-loop renormalisation of helicity matrix elements is shown
calculated with a dipole form factor with varying values of the $\pi
N\Delta$ coupling, $g_{\pi N \Delta}$. Full details are given in
Ref.~\cite{DMT2}. For these helicity (and transversity) operator
matrix elements the difference between $g_{\pi N \Delta}/g_{\pi NN}=0$
(no $\Delta$) and $g_{\pi N \Delta}/g_{\pi NN}=2$ (phenomenological
value) is up to 15\% --- much greater than in the unpolarised case
where the overall effect is less than 2\% over the entire mass range
studied here.

In order to incorporate these effects into the extrapolations of
lattice data, we calculate the required renormalisations for the
phenomenologically preferred dipole parameter, $\Lambda=0.8$~GeV, and
for varying values of the coupling ratio $g_{\pi N\Delta}/g_{\pi NN}$.
We then determine $\mu$ by fitting to the calculated renormalisations
using Eq.~(\ref{xtrap}) (with $c_{\rm LNA}$ calculated in the $\Delta
M= M_\Delta -M_N \to 0$ limit). With $\mu$ thus fixed, we then use the
lattice data to determine the fit parameters $a_n$ and $b_n$. The
resulting curves are then shown in each panel of Fig.~\ref{f2} for
$g_{\pi N\Delta}/g_{\pi NN}=0$ (no $\Delta$), $\sqrt{72/25}$, 1.85 and
2.\footnote{Reasonable variation of the strength of the
  Weinberg-Tomozawa contact term is also included -- see
  Ref.~\cite{DMT2} for details.}

\begin{center}
  \begin{table}
    \begin{tabular}{ccc}
    Moment& Experimental & Extrapolated \\
    \hline 
    $\langle x \rangle_{u- d}$ & 0.145(4) & 0.17(3)\\ 
    $\langle x^2 \rangle_{u- d}$ & 0.054(1)& 0.05(2)\\ 
    $\langle x^3 \rangle_{u- d}$ & 0.022(1) & 0.02(1)\\
    $\langle 1 \rangle_{\Delta u- \Delta d}$ & 1.267(4) & 1.12(8) \\
    $\langle x \rangle_{\Delta u- \Delta d}$ & 0.210(25) & 0.27(3) \\ 
    $\langle x^2 \rangle_{\Delta u- \Delta d}$ & 0.070(11) &0.14(4) \\  
    $\langle 1 \rangle_{\delta u- \delta d}$ & ? & 1.22(8) \\ 
    $\langle x \rangle_{\delta u- \delta d}$ & ? & 0.5(1) \\ 
  \end{tabular}
  \caption{\label{t1} Values of the unpolarized, helicity and 
    transversity moments, extrapolated to the physical pion mass 
    using Eq.~(\ref{xtrap}). For comparison, experimental values 
    of the moments, where known \cite{param}, are also listed.}
\end{table}
\end{center}

\vspace{-10mm} Table \ref{t1} shows the resulting extrapolated values
along with uncertainties resulting from the $\pi N\Delta$ and
Weinberg-Tomozawa couplings, statistical and (estimated) systematic
errors. Also listed are the experimental moments where known
\cite{param}.  It is evident that the agreement between the
experimental and extrapolated results is very good for the unpolarised
moments, but significantly worse in the helicity sector.  The
extrapolated value of $g_A$ is 10\% below its experimental value, with
8\% errors.  However, there is some evidence that lattice simulations
of this quantity are particularly susceptible to finite volume effects
\cite{Sasaki:2001th}. For the higher helicity moments, statistical
errors on the lattice data are not yet sufficiently small to make
definitive statements.  Since the renormalisation of the transversity
matrix elements is almost identical to that of the helicity matrix
elements, we are also able to make an estimate of the values of these
moments which can be compared with future experimental determinations.

\section{Conclusion}

We have investigated the extrapolation of lattice data on the low
moments of the isovector, unpolarised, helicity and transversity
parton distribution functions. In the case of the polarised moments,
the $\Delta$ isobar, even though not leading non-analytic, was found
to play a significant role and its inclusion in the operator
renormalisation and extrapolation procedure is vital. With this
accounted for, we have the surprising result that the effect of the
non-analytic behavior is strongly suppressed for the polarised
moments, and a naive linear extrapolation of the moments provides
quite a good approximation to the more accurate form.

Talk given by W. D. who is grateful to R.  Alkofer and the
Institut f\"ur Theoretische Physik, Universit\"at T\"ubingen for their
hospitality and to DFG grant Al279/3-3 for support. W.M. is supported
by the U.S. Department of Energy contract \mbox{DE-AC05-84ER40150},
under which the Southeastern Universities Research Association (SURA)
operates the Thomas Jefferson National Accelerator Facility (Jefferson
Lab).

\end{document}